\documentstyle[prb,aps,psfig]{revtex}
\begin{document}
\draft

\title{Finite temperature molecular dynamics study of 
unstable stacking fault free energies in silicon}
\author{M. de Koning and A. Antonelli}
\address{Instituto de F\'{\i}sica Gleb Wataghin,
 Universidade Estadual de Campinas, \\ Unicamp 
 13083-970, Campinas, S\~ao Paulo, Brazil}
\author{Martin Z. Bazant}
\address{Department of Mathematics,\\
Massachusetts Institute of Technology, Cambridge, MA 02139}
\author{Efthimios Kaxiras}
\address{Department of Physics and Division of Applied Sciences,\\
Harvard University, Cambridge, MA 02138}
\author{J.F. Justo}
\address{Instituto de F\'{\i}sica da Universidade de S\~ao Paulo, \\
CP 66318, CEP 05315-970    S\~ao Paulo -  SP, Brazil}
%
\maketitle

\begin{abstract}
\tightenlines
We calculate the free energies of unstable stacking fault (USF) 
configurations on the glide and shuffle slip planes in silicon 
as a function of temperature, using the recently developed 
Environment Dependent Interatomic Potential (EDIP). 
We employ the molecular dynamics (MD) adiabatic switching method
with appropriate periodic boundary conditions and restrictions 
to atomic motion that guarantee stability and include 
volume relaxation of the USF configurations perpendicular to
the slip plane. Our MD results using the EDIP
model agree fairly well with earlier first-principles estimates for 
the transition from shuffle to glide plane dominance as a function
of temperature. We use these results to make contact to  
brittle-ductile transition models. 
\end{abstract}

%
%
The physics describing the behavior of extended defects at the microscopic 
level and its relation to macroscopic mechanical properties of materials, 
which are of fundamental importance to technology, 
have been subject of investigation for many years.  An example 
of intriguing macroscopic behavior is   
the brittle-ductile transition (BDT),  corresponding to a change 
in the state of the system from a brittle, easily fractured, 
into a ductile, tough substance that can easily undergo plastic deformation. 
Silicon is a material in which the BDT is particularly spectacular,
taking place over a very narrow temperature range of only a few degrees $K$,
\cite{Samuels,Hirsch} at a critical temperature near 873 $K$. 
Brittle or ductile behavior is related to the response of a sharp 
crack tip to external loading: Brittleness is typically associated 
with easy crack propagation, whereas ductility is characterized by 
blunting of the crack tip through the emission of dislocations. 
The microscopic mechanisms behind the crack tip response are 
related to the ability of the solid to nucleate and emit 
dislocations from the loaded crack tip.\cite{Hirth} 

While complex atomistic processes such as dislocation nucleation 
and mobility cannot easily be captured by simple
phenomenological models, certain features of the structure of Si may be
related to the abruptness of the BDT in this material.
In silicon two distinct sets of closely packed \{111\} slip
planes, called ``glide'' and ``shuffle'' sets\cite{Hirth}, 
are relevant for dislocation nucleation and slip. 
Dislocations nucleated on the shuffle set are relatively narrow making 
the resistance to dislocation motion, the so-called Peierls stress, 
relatively high and their mobility low. 
On the glide set the mobility properties are different. 
Due to the splitting of dislocations into partials on this set, 
the corresponding Peierls stresses are smaller and the dislocation 
mobility is higher. In this sense, the abruptness of the BDT in 
silicon may be associated with a sudden change in the dominance  
of one set over the other.
In order to 
characterize such a change, a theoretical model is required that
can link the processes of dislocation nucleation and motion to 
simple, material-specific quantities which can be calculated accurately. 

Recent theoretical work by Rice and collaborators 
\cite{Rice,Rice-Beltz,Sun-Beltz-Rice} 
developed such a model of dislocation nucleation at a crack tip 
based on a continuum elasticity approach and the 
Peierls stress concept.  From  
this analysis, the so-called unstable stacking energy $\gamma_{us}$ 
was found to be a measure for the resistance to dislocation nucleation 
at a crack tip (we refer to this below as the dislocation nucleation
criterion). The unstable stacking energy, which corresponds to
the unstable stacking fault (USF) configuration,\cite{Kaxiras,Juan}
is defined as the 
lowest energy barrier that needs to be crossed when one half of 
a perfect crystal slips, on a dislocation nucleation plane of interest, 
relative to the other half, completing a total displacement equal to 
one lattice repeat vector.
 
The calculations performed by 
Kaxi\-ras and Dues\-bery\cite{Kaxiras} and Juan and Kaxi\-ras\cite{Juan}, 
based on first principles Density Functional Theory techniques 
within the Local Density Approximation (DFT-LDA), provided accurate 
values for $\gamma_{us}$ on the shuffle and glide 
planes of Si. Through the use of Vineyard's Transition State 
Theory (TST),\cite{Vineyard} these zero temperature results were extended 
to finite temperature and pressure conditions.\cite{Kaxiras,Juan}
This approach, while illustrating the basic idea of an abrupt transition 
from shuffle to glide set dominance as a function of temperature, 
is theoretically limited, because it neglects real dynamics of atoms 
on either side of the slip plane and it relies on mapping the
system to an oversimplified two-dimensional model. 
It is therefore desirable to include finite temperature 
effects in an explicit manner. To this end, 
finite temperature molecular dynamics (MD) simulation of the 
USF configurations would represent an interesting improvement. Within such a
context the application of a finite temperature first-principles MD method
would be most appropriate. However, as long as the present computational 
limitations of such techniques inhibit their application to larger systems over
wide temperature ranges, an approximate approach based on empirical models remains the only alternative.
With the development of the
novel empirical Environment Dependent Interatomic Potential (EDIP) for 
silicon,\cite{Bazant,Justo} the application of such an approach seems now 
feasible. As opposed to other empirical
models,\cite{Stillinger,Tersoff} EDIP captures with adequate 
realism several important stable, metastable and saddle-point 
configurations, including the
energetics of generalized stacking faults and dislocation cores, and
promises to be very useful for the description of the dynamics of USF
configurations. 

In this brief report 
we address the question of the influence
of finite temperature effects on the dislocation nucleation criterion, 
through an explicit calculation of the free energies associated with the USF
configurations on the glide and
shuffle sets as a function of temperature. For this purpose 
we employ the MD adiabatic switching method which is based on the
simulation of thermodynamically reversible processes and enables the
efficient and quantitatively reliable determination of
thermal quantities including all anharmonic effects.
\cite{Watanabe,de Koning1,de Koning2} We use the EDIP model for the 
description of the interatomic forces in the 
MD simulations and compare our results to DFT-LDA 
results.\cite{Juan}

Two specific technical issues are involved in the MD simulation of USF 
configurations.
The first concerns the definition of the two atomic USF configurations 
(shuffle and glide) within
MD cells. For this purpose, the choice of an orthogonal
coordinate system formed by axes parallel to $[111]$, $[10\bar{1}]$, and
$[1\bar{2}1]$ directions is most appropriate.
The glide and shuffle USF configurations are defined
by the slip displacement vectors $\frac{1}{12} [1\bar{2}1]$ and $\frac{1}{4}
[10\bar{1}]$
(in units of the lattice parameter)\cite{Kaxiras,Juan} respectively, 
which describe the relative displacement of the two atomic blocks adjacent to 
the crystal slip plane under consideration. These properties are implemented
by adding the corresponding slip displacement vectors on the periodic
boundary repeat vectors in the $[111]$ direction of the computational cells.

The second issue is related to the fact that the USF configurations are 
intrinsically unstable. Due to the state of slip of these configurations,
considerable shear
stresses appear in the system, which tend to relax through shear strain.
In order to prevent
this relaxation during the MD simulations, the motion of
the atoms in the two planes immediately adjacent to the slip planes is
restricted to the $[111]$ direction, perpendicular to the shear stresses.

The size of the computational cells utilized for the simulation of the USF 
configurations is choosen such that the interaction between the periodic 
images is negligibly small. To this end, the number of atoms for both the
glide and shuffle USF configurations was fixed at 648, divided in 36 atomic 
(111) planes. In order to allow for volume relaxation perpendicular to the 
slip planes, $\gamma_{us}$ (including atomic relaxation) is evaluated at 
several volumes below and above the ideal volume of bulk silicon. 
For the glide set, the minimum value of $\gamma_{us}$ 
is obtained for a perpendicular expansion $\Delta z_{glide}=0.35 \, \AA$, 
while the minimum value of $\gamma_{us}$ for 
the shuffle set USF configuration is obtained for a contraction  
$\Delta z_{shuffle}=-0.22 \, \AA$.

\begin{table}
\vspace{0.2in}
\begin{center}
\begin{minipage}[h]{5.5in}
\caption{Static (0 $K$) unstable stacking energy $\gamma_{us}$ for the \{111\}
shuffle and glide set in silicon at various levels of relaxation. All values are
in $\mbox{J\,m}^{-2}$. The DFT results were taken from Juan and Kaxiras (Ref.
6).}
\vspace{0.2in}
\begin{tabular}{c|c|c}
& Shuffle set \ \ \ \  &  Glide set \ \ \ \  \\
\hline
No relaxation && \\
\hline
DFT-LDA & 1.84 \ \ \ \ & 2.51 \ \ \ \ \\
EDIP & 1.98 \ \ \ \ & 3.28 \ \ \ \ \\
\hline
Atomic relaxation &&\\
\hline
DFT-LDA & 1.81 \ \ \ \ & 2.02 \ \ \ \ \\
EDIP & 1.28 \ \ \ \ & 1.89 \ \ \ \ \\
\hline
Atomic+volume relaxation && \\
\hline
DFT-LDA & 1.67 \ \ \ \ & 1.91 \ \ \ \ \\
EDIP & 1.04 \ \ \ \ & 1.86 \ \ \ \ \\
\end{tabular}
\label{table3}
\end{minipage}
\end{center}
\end{table}

Table \ref{table3} shows several 0 $K$ $\gamma_{us}$ values characterized
by different levels of relaxation, as obtained with EDIP and
DFT-LDA.\cite{Juan} Although EDIP correctly predicts $\gamma_{us}$ 
to be higher for the glide set than for the shuffle set, the quantitative
discrepancies with DFT-LDA are significant. The consequences of these 
discrepancies will be discussed further below. 
Despite the quantitative differences, EDIP successfully captures
the qualitative trends predicted by DFT-LDA for the effects of
relaxation on $\gamma_{us}$.
Both approaches predict the influence of atomic relaxation to
be larger on the glide set, and that 
volume relaxation is more pronounced on the shuffle set.

In order to evaluate the free energies of the USF configurations, a
series of MD adiabatic switching simulations is performed for several
temperatures between 200 and 1600 $K$.
In these simulations, the interacting silicon atoms are
transformed into identical harmonic 
oscillators, under $(N,V,T)$ conditions.\cite{de Koning1,de Koning2} The 
equilibrium positions of the oscillators are centered at the equilibrium 
positions of the silicon atoms in
the USF configurations and all oscillators have the same characteristic
frequency. Effects of thermal expansion are taken into 
account using lattice constants determined from standard $(N,P,T)$ simulations
for bulk silicon.

\begin{figure}[b]
\vspace{0.5in}
\centering{
\mbox{\psfig{figure=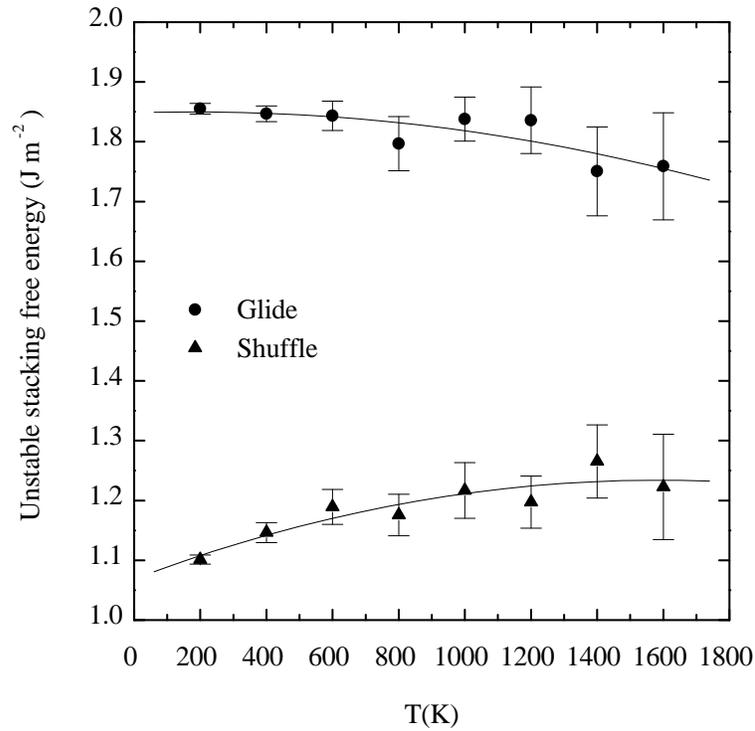,height=4in}}
\begin{minipage}[h]{5.5in}
\vspace{0.3in}
\caption{USF Helmholtz free energies
as a function of temperature,
for the glide and shuffle sets in silicon at zero pressure.
}
\label{bdt2}
\end{minipage}
}
\end{figure}

\begin{figure}
\begin{center}
\mbox{
\psfig{figure=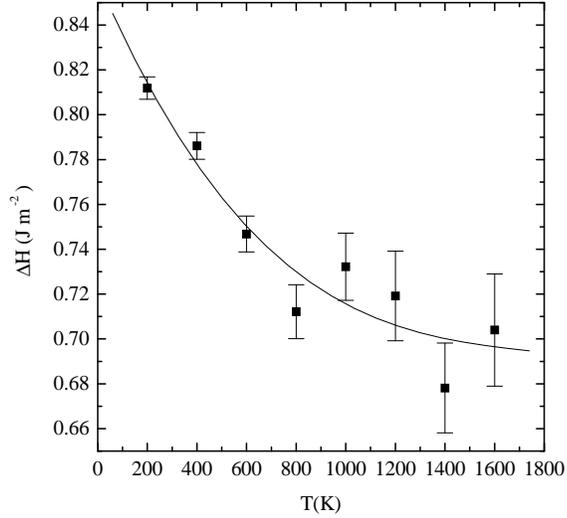,height=3in}
}
\vspace{0.1in}
\begin{minipage}[h]{5.5in}
\caption{Zero pressure enthalpy difference
between the glide and shuffle USF configurations
as a function of temperature.
}
\label{bdt1}
\end{minipage}
\end{center}
\end{figure}

\begin{figure}
\begin{center}
\mbox{
\psfig{figure=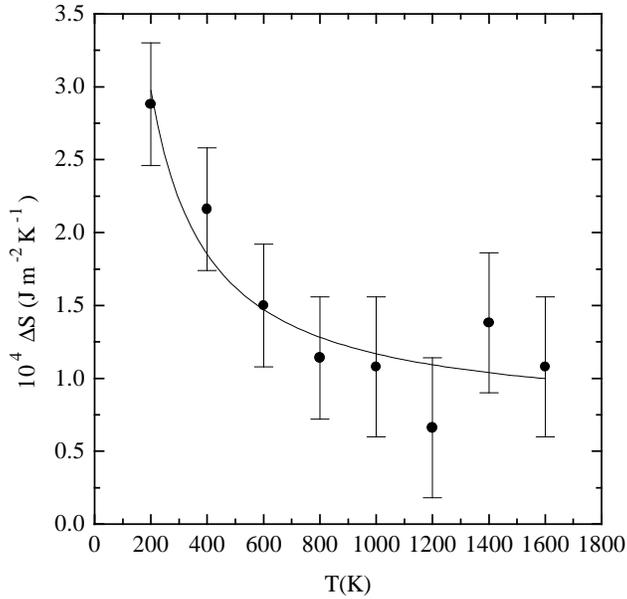,height=3.3in}
}
\vspace{0.1in}
\begin{minipage}[h]{5.5in}
\caption{Entropy difference between the glide and shuffle USF
configurations as a function of temperature.}
\label{bdt3}
\end{minipage}
\end{center}
\end{figure}

The results of these calculations are summarized in figure \ref{bdt2}, which 
shows the USF configuration
Helmholtz free energies at zero pressure 
for the
glide and shuffle sets as a function of temperature.
Over the entire temperature range, the 
Helmholtz free energy of the glide set USF configuration is higher than
for the shuffle set, although the difference
between them decreases with increasing
temperature. Figures \ref{bdt1} and \ref{bdt3} show 
the respective contributions of the enthalpy and entropy
differences between 
the glide and shuffle USF configurations. 
 
At this point it is interesting to analyze the results of these MD 
simulations within the framework of Rice's theory on dislocation nucleation.
So far, the calculations have ignored
the possible influence of finite pressure conditions. These effects should
be taken into account in order to obtain a more realistic
picture of the energetics involved in the dislocation nucleation criterion.
To this end the introduction of the Gibbs free energy per unit area
according to the definition adopted by Kaxiras and Duesbery\cite{Kaxiras} and 
Juan and Kaxiras\cite{Juan} is appropriate:
\begin{equation}
G(P,T)=F(T)+P \Delta z.
\end{equation} 
Here $F(T)$ is the Helmholtz free energy per unit area (determined from the 
adiabatic switching simulations) and $\Delta z$ represents the
volume relaxation perpendicular to the slip plane under consideration. 
According to the dislocation nucleation criterion, the condition that the
preferred slip plane changes from shuffle to glide is given by
$G_{shuffle}(P,T)=G_{glide}(P,T)$, which describes the $(P,T)$ coexistence
curve separating the two dislocation nucleation regimes in the phase diagram.

Before embarking on the construction of the EDIP phase diagram, the question 
related to
the quantitative $\gamma_{us}$ discrepancy mentioned earlier should
be addressed.  
According to Table \ref{table3}, EDIP overestimates the difference between the 
glide and shuffle values of $\gamma_{us}$ at zero temperature (including atomic 
and volume
relaxation) by more than a factor three. Since this discrepancy is reflected in 
the finite temperature values of
$\gamma_{us}$, it strongly affects the free energies and distorts the
corresponding EDIP phase diagram. In order to eliminate this effect, the EDIP
Helmholtz free energy differences are corrected in such a manner that the zero 
temperature value equals the corresponding DFT-LDA value. This correction is 
most conveniently
accomplished by means of a simple rigid energy shift imposed for all 
temperatures. Such a shift only modifies the static energy scales while it
leaves unaltered the dynamical entropic properties.  

\begin{figure}[b]
\vspace{.8in}
\begin{center}
\mbox{
\psfig{figure=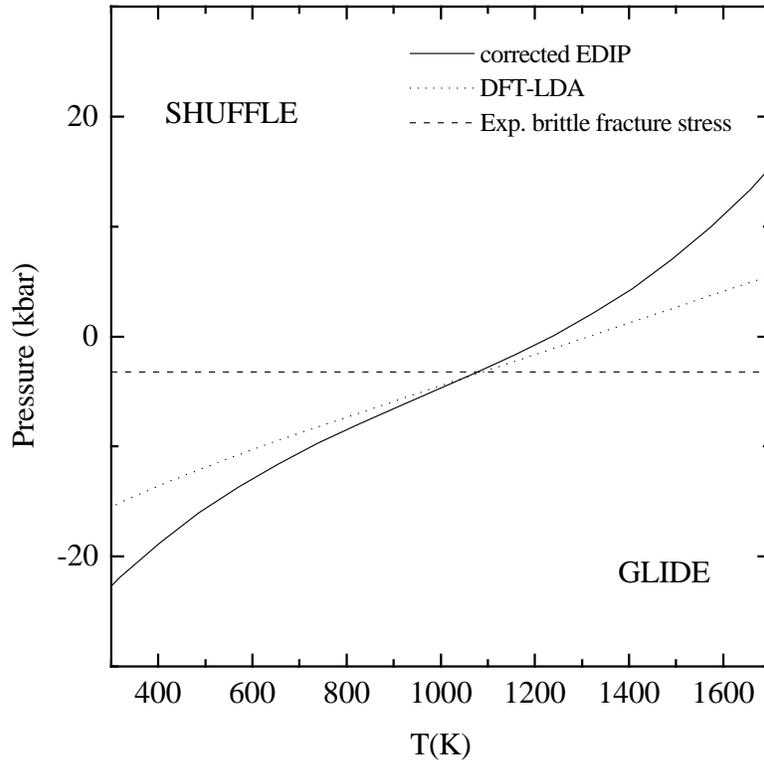,height=4in}
}
\vspace{0.2in}
\begin{minipage}[h]{5.5in}
\caption{Phase diagram: Coexistence curves separating the preferable nucleation
of dislocations on shuffle planes versus glide set. The full line
represents the results obtained with the EDIP potential after correction for the
overestimated difference in $\gamma_{us}$. The dotted line represents the
DFT-LDA calculations by Juan and Kaxiras (Ref. 6). The dashed line represents
the experimental brittle fracture stress.}
\label{bdt5}
\end{minipage}
\end{center}
\end{figure}

Figure \ref{bdt5} shows a comparison between the corrected EDIP phase diagram 
(continuous line)
and the DFT-LDA results\cite{Juan} (dotted line). For $(P,T)$ values below
(above) these curves, the glide (shuffle) set USF configuration has lower
free energy. Both coexistence curves agree fairly well, with
derivatives $dP/dT$ of the same order of magnitude over the entire temperature 
range. Furthermore, the curves intersect near 1100 $K$ with nearly equal slopes.
The most significant difference between both curves is the inflection point which appears in the EDIP phase diagram and originates from the
specific behavior of the glide-shuffle
entropy difference as a function of temperature (negative first
derivative and positive second derivative).
However, despite this discrepancy, both phase diagrams show no pronounced
differences. This 
suggests that the exact treatment of temperature dependent
vibrational effects in the present simulations does not significantly alter
the qualitative picture resulting from the approximate TST approach adopted in 
earlier DFT-LDA calculations. 

The phase diagrams in figure \ref{bdt5} illustrate the basic idea of an abrupt 
transition from shuffle to glide set dominance under specific temperature and 
pressure conditions. Such a transition may be related to the sharp BDT
transition observed in silicon. Although dislocations nucleate more easily
on shuffle planes, the splitted Shockley partials on the glide set are more
mobile. In this sense, the transition from brittle to ductile behavior
might be directly related to the abrupt transition from shuffle to glide set
dominance. Within such an interpretation, the EDIP and DFT-LDA phase diagrams 
might indicate a transition temperature at the intersection
points of the coexistence curves with the experimental brittle fracture stress.
In this manner both EDIP and DFT-LDA would predict a transition temperature
of 1100 $K$ which is not unreasonably far from 
the experimental critical temperature
of 873 $K$. 

These results can not be taken literally 
for several reasons however. 
First, important factors such as 
electronic entropy contributions, surface free energies and dislocation core 
reconstruction effects have been neglected in the present approach.
Furthermore, the microscopic
mechanisms involved in the BDT transition phenomenon may involve subtleties  
which are not included in the approach adopted in this work. Therefore, a 
direct quantitative interpretation of the dislocation nucleation criterion
represents an oversimplification, although we expect the qualitative
picture furnished by this approach to be reasonable. 

In summary, we have calculated the free energies of the shuffle and glide USF 
configurations in silicon as a function of temperature. For this purpose we
used the recently developed empirical EDIP model and applied the MD adiabatic
switching method which allows accurate and efficient determination of
thermal quantities including all anharmonic effects. The results of the
finite temperature MD simulations agree fairly well with
earlier DFT-LDA calculations on the transition from shuffle to glide dominance as a function of temperature. 
This suggests that the full inclusion of finite temperature effects in our simulations does not significantly alter the qualitative
picture provided by the
TST DFT-LDA approach, in which such effects were treated in an approximate 
manner.

M.K. and A.A. acknowledge financial support from FAPESP, CAPES, FAEP,
and CNPq,
and J.F.J. acknowledges support from FAPESP, all of 
them Brazilian funding agencies. 
M.B. and E.K. acknowledge support by the Harvard MRSEC and J.F.J.  
acknowledges support by the MIT MRSEC, both of which are funded 
through NSF.


\end{document}